\documentclass{elsart}
\usepackage{epsfig}
\journal{Physics Letters {\bf B}}
\newcommand{\beq}  {\begin{equation}}
\newcommand{\eeq}  {\end{equation}}
\newcommand{\bmath}{\begin{eqnarray}}
\newcommand{\emath}{\end{eqnarray}}
\newcommand{\bei}{\begin{itemize}}
\newcommand{\eei}{\end{itemize}}
\newcommand{\mgg}{m_{\gamma \gamma}}

\newcommand{\klpgg}{K_{L}\rightarrow\pi^0\gamma\gamma}
\newcommand{\klpee}{K_{L}\rightarrow\pi^0 e^+e^-}
\newcommand{\kspp}{K_{S}\rightarrow{\pi^0\pi^0}}
\newcommand{\klpp}{K_{L}\rightarrow{\pi^0\pi^0}}
\newcommand{\klppp}{K_{L}\rightarrow{\pi^0\pi^0\pi^0}}

\def\pgg{\pi^0\gamma\gamma}
\def\mgg{m_{\gamma\gamma}}

\begin{document}
\hyphenation{cha-rac-te-ri-zes}
\begin{frontmatter}
\title{\bf Precise measurement of the decay {\boldmath $\klpgg$}}
\date{\today}
\author{A.~Lai},
 \author{D.~Marras}
\address{Dipartimento di Fisica dell'Universit\`a e Sezione dell'INFN di Cagliari, I-09100 Cagliari, Italy.} 
\author{A.~Bevan},
 \author{R.S.~Dosanjh},
 \author{T.J.~Gershon\thanksref{thref1}},
\thanks[thref1]{ Present address: KEK, Tsukuba, Ibaraki, 305-0801, Japan.}
\author{B.~Hay\thanksref{thref2}},
\thanks[thref2]{ Present address: EP Division, CERN, 1211 Gen\`eve 23, Switzerland.}
\author{G.E.~Kalmus\thanksref{thref3}},
\thanks[thref3]{Based at Rutherford Appleton Laboratory, Chilton,
 Didcot, OX11 0QX, U.K.}
 \author{C.~Lazzeroni},
 \author{D.J.~Munday},
\author{M.D.~Needham\thanksref{thref4}},
\thanks[thref4]{ Present address: NIKHEF, PO Box 41882, 1009 DB
 Amsterdam, The Netherlands.}
\author{E.~Olaiya},
 \author{M.A.~Parker},
 \author{T.O.~White},
 \author{S.A.~Wotton}
\address{Cavendish Laboratory, University of Cambridge, Cambridge, CB3 0HE, U.K.\thanksref{thref5}}
\thanks[thref5]{ Funded by the U.K.    Particle Physics and Astronomy Research Council.}
% ...
\begin{abstract}
The decay rate of $\klpgg$ has been measured with the NA48 detector at 
the CERN SPS. A total of $2558$ $\klpgg$ candidates have been observed
with a residual background of $3.2\%$.
The branching ratio is determined to be $(1.36 \pm0.03_{(stat)} 
\pm0.03_{(syst)}\pm 0.03_{(norm)})\times10^{-6}$ and the vector coupling constant 
$a_v = -0.46 \pm0.03_{(stat)} \pm0.04_{(syst)}$. This result suggests
that the CP-violation effects are dominating in the $\klpee$ decay.
An upper limit for the $\klpgg$ decay rate in the two photon mass region
$m_{\gamma\gamma}<m_{\pi^0}$ is also given. 
\end{abstract}
\end{frontmatter}

\section{\bf Introduction}

The measurement of the $\klpgg$ decay is useful to constrain the CP conserving 
amplitude of the decay $K_{L}\rightarrow\pi^0 e^+ e^-$ via two photon
exchange. 
Previous measurements of the decay \cite{na31-1,na31-2,e731,ktev} 
have been compared with calculations performed in the framework of the 
Chiral Perturbation Theory ($\chi PT$),
the effective theory of the Standard Model at low energy in the
hadronic sector. These predictions are best described in terms of the
two amplitudes A and B (referring to angular momentum states $J=0$ and 
$J=2$ of the two photons respectively) in the Lorentz invariant
expression for the double differential decay rate \cite{D'Ambrosio}:
\beq
\frac{\partial^2\Gamma}{\partial y \partial z} = \frac{m_K}{2^9 \pi^3}
\left[ z^2\cdot|A+B|^2+\left(y^2- y_{\rm max}^2\right)^2\cdot |B|^2\right]
\label{eq:ampli} 
\eeq
$$z=\frac{(k_3+k_4)^2}{m_K^2}=\frac{m_{3,4}^2}{m_K^2} ~~~~~~~~~~~~~
y=\frac{|p_K(k_3-k_4)|}{m_K^2}$$
with $m_K$ and $p_K$ the kaon mass and momentum, $k_3$ and $k_4$ the
two photon momenta and $y_{\rm max}$
the kinematic bound for the $y$ variable, given by:
$$ y_{\rm max} = \frac{1}{2}
\sqrt{\left(1-\frac{{m_{\pi}}^2}{{{m_K}^2}}\right)^2+
  \left(1+\frac{{m_{\pi}}^2}{{{m_K}^2}}\right)^2\times z+z^2}
$$ 
The leading $O(p^4)$~~$\chi PT$ calculation predicts B=0,
in qualitative agreement with the experimental observation of a $z$
spectrum peaked at high values, but underestimates the $\klpgg$
branching ratio by about a factor of three \cite{Ecker1}. 
At $O(p^6)$ the rate and the $m_{\gamma\gamma}$ spectrum can be 
reproduced by adding a contribution from the VMD 
mechanism \cite{D'Ambrosio,Ecker2,Sehgal}, via the coupling constant 
$a_v$ \cite{Ecker3,D'Ambrosio0,Kambor}. 
The VMD mechanism could enhance the state $J=2$ for the two photons, 
hence allowing a sizeable  CP conserving contribution which is
not helicity suppressed  for the $K_{L}\rightarrow\pi^0 e^+ e^-$ decay \cite{Donoghue}.\\
In this paper we present a new measurement
of the branching ratio and decay spectrum of the decay $\klpgg$, 
based on data collected
in the years 1998 and 1999 with the apparatus of the NA48 experiment.
Quantitative information on the VMD contribution is obtained  
from the spectrum at low z values.

\section{\bf Experimental set-up}

The $\klpgg$  events are collected using the NA48 detector located at the 
CERN-SPS and primarily devoted to measure the parameter 
$\epsilon^{\prime}/\epsilon$ which characterises direct CP violation 
in neutral Kaon decays into two pions \cite{na48}\cite{na48eprime9899}.
Neutral Kaons are produced in interactions of  $1.4\times 10^{12}$~450~GeV/$c$ 
protons on a beryllium target during 2.4~seconds every 14.4~seconds. 
A system of sweeping magnets and collimators defines a 
neutral beam of $2\times 10^{7} K_L$ per burst and 
divergence of $\pm 0.15$~mrad.
A fraction of the non interacting protons are redirected by channelling
in a bent crystal, to a second target, 120~metres downstream of the
first, to generate a $K_S$ beam \cite{doble}.
In order to tag $K_S$ decays, the protons before hitting this target are
detected  by an array of scintillation counters (tagger).
The two beams converge with an angle of $0.6$~mrad at the calorimeter, 
$\approx$ 120~metres downstream of the final collimators.
The decay region is contained in an evacuated cylindrical vessel 89 metres 
long separated from the NA48 detector volume by a  thin
Kevlar window $0.3\%$ of a radiation length ($X_0$).
The neutral beam traverses the detector inside a 16~cm diameter vacuum pipe. 
A scintillation counter (AKS) placed on the $K_S$ beam defines the 
upstream edge of the fiducial region for decays from that beam.\\
The most important detector element for this analysis is the quasi-homogeneous
liquid Krypton electro-magnetic calorimeter (LKr) structured in 13212 readout 
tower cells $27~X_0$ deep \cite{gunal}. The ionization signal from each of 
the cells is integrated, amplified, shaped and digitised by 10-bit FADCs at 
40~MHz sampling frequency.
The energy resolution can be parameterised as:
$$ \sigma(E)/E \simeq (0.09\pm 0.01)/E \oplus (0.032\pm
0.002)/\sqrt{E} \oplus (0.0042\pm 0.0005), \nonumber $$
with E in GeV. The spatial and time resolutions are better than $1.3$~mm and $300$~ps, 
respectively, for photons with energy greater than 20~GeV.\\
Seven ring shaped counters (AKL), consisting of plastic scintillator and iron 
converters, surround the decay and spectrometer region in order to veto events 
with photons outside the calorimeter acceptance. Their efficiency has been
estimated to be around $95\%$.
The charged particles are reconstructed by a magnetic spectrometer consisting of 
four drift chambers (DCH) and a magnetic dipole. 
The space resolution for each projection
is $90~\mu$m and the average efficiency is better than $99.5\%$ per plane. For 
this analysis the spectrometer is used to veto events with
reconstructed tracks.
  
\section{\bf Data taking}

Events of the $\klpgg$ decay channel and of the channel
$\klpp$, used for normalisation, are collected by the same
trigger, since their final states appear identical in all 
detector elements.
The trigger decision is based on quantities which are 
derived from the sums of the energy deposited in the LKr calorimeter
in groups of cells corresponding to 64 horizontal and 64 vertical 
slices \cite{fischer}. 
The trigger requirements are:
\bei
\item[1)] at maximum 5 peaks in each projection;
\item[2)] total energy larger than 50~GeV;
\item[3)] centre of gravity of the event, computed 
from the first moments of the energy peaks in the projections to be 
within 15~cm from the beam axis;
\item[4)] proper time of the decay, computed from the second moments, 
less than $5~\tau_{K_S}$ from the beginning of the decay region.
\eei
The trigger efficiency has been checked to be better than  $99.9\%$ \cite{na48eprime9899} 
using  a minimum bias sample triggered by a scintillating fibre 
hodoscope placed in the LKr volume at a depth of $9.5~X_0$. \\
During the experimental runs roughly 180 Terabytes of raw data were recorded 
and pre-selected by an on-line software filter, whose criteria are 
subsequently tightened in the off-line analysis.

\section{\bf Analysis }
The first step is to reconstruct the showers in the liquid krypton 
calorimeter summing the energies of the cells in a circle of $11$~cm radius 
from a local maximum.
The time and the position of the shower are derived from the most energetic 
cells and from the centre of gravity of the energy. A partial overlap between two 
showers is resolved using the shower profile obtained from simulation. The 
shower energy is corrected for:
\bei
\item[i)] the residual 0.4\% non-uniformity in the calibration electronics chain, 
comparing energy and momentum of electrons from $K_{e_3}$ events as a
function of the hit cell. The uniformity is improved to the 0.15\% level;
\item[ii)] the 1\% peak-to-peak modulation of the response as a
  function of the impact point 
in the LKr cell due to the effects of a $2$~mm gap between electrodes
in the vertical direction  and to the finite integration time
in the horizontal one;
\item[iii)] the energy-leakage in the beam pipe at small radius, the
  material before the active volume and the presence of 
  not properly working cells;
\item[iv)] the small (0.1\%) non-linearity. 
\eei
Since the $\klpgg$ event is characterised by four clusters with the only 
constraint of two showers coming from a $\pi^0$ decay,
all the events with at least four clusters are considered.
Each cluster must be located well inside the calorimeter 
volume, i.e. within an octagon of 113~cm apothem, with 15~cm inner radius.
It must be at a distance larger than  2~cm from any non working cell 
and separated from another cluster by 
at least $10$~cm. The  cluster energy must be between 3~GeV and 
100~GeV and the time within  3~ns of 
the event time, defined as the average of the cluster times. 
The total energy of the clusters must be in the 70-170~GeV interval.\\
The two photon invariant mass is:
\beq
m_{\gamma_i \gamma_j}= \frac{\sqrt{{E_i}{E_j}d_{i,j}^2}}{z_{LKr}- z_K}
\label{mgg}
\eeq
where $E_i, E_j$ are the cluster energies, $d_{i,j}^2=[(x_i - x_j)^2 + (y_i -
y_j)^2]$ with $x_i$ and $y_i$ the 
horizontal and vertical positions of the i-th cluster at the LKr front
face.
The longitudinal coordinate of the decay vertex, $z_K$, is computed
from the electro-magnetic calorimeter 
information assuming the nominal kaon mass for the parent particle:
\beq       
z_{K}= z_{LKr} - \frac{\sqrt{\sum_{i,j,i>j}{E_i}{E_j}d_{i,j}^2}}{m_K}, 
\label{zk}
\eeq
where $z_{LKr}$ is the longitudinal coordinate of the LKr front face.
The overall energy scale is fixed by the fit of the position of AKS
counter, that vetoes all $K_S$ decays occurring upstream \cite{na48}.\\
We only consider events for which there is no proton detected in the
$K_S$ tagger in time (within $\pm 2$~ns) with the event; 
this rejects $10.65\%$ \cite{na48eprime9899} of the $K_L$
decays, due to accidental coincidences with a proton in the tagger.\\
Because of the small branching ratio, the $\klpgg$ signal is subject to large 
backgrounds such as $\klppp$ decays, with missing and/or overlapping photons, 
mis-reconstructed $\klpp$ decays with $\gamma$ conversion in the 
spectrometer and events generated by accidental pile up of particles
from two different events.
We distinguish two classes of cuts in the analysis: 
$2\pi^0$ rejection and  $3\pi^0$ rejection.\\
\subsection{\bf $2\pi^0$ rejection.}
Three different combinations of ${\gamma \gamma}$ pairs can be formed from 
four showers. A $\chi^2$ variable is defined: 
\beq
 \chi^2=\left[ \frac{(m_{1,2}+m_{3,4})/2-m_{\pi^0}}{\sigma_+}\right]^2+
        \left[ \frac{(m_{1,2}-m_{3,4})/2}{\sigma_-}\right]^2
\eeq
where $\sigma_{\pm}$ are the resolutions of $(m_{1,2}\pm m_{3,4})/2$ measured
from the data and parametrised as a function of the lowest photon energy, 
typically $\sigma_+\approx 0.42$~MeV/$c^2$ and 
$\sigma_-\approx 0.83$~MeV/$c^2$. 
In order to select a $\pi^0\pi^0$ candidate the best combination is 
required to have $\chi^2 \leq 13.5$, while the $\pi^0\gamma\gamma$
events are selected with $\chi^2 > 300$.
We retain for further analysis all combinations that give a photon
pair (that we call $m_{1,2}$) within 3~MeV/$c^2$ from $m_{\pi^0}$ 
and the other pair (corresponding to $m_{3,4}$)  outside the 
window $110-160$~MeV/$c^2$.\\
The $2\pi^0$  background  concentrates in the most interesting region
for the determination of $a_v$, namely at low 
$m_{3,4}$ value ($m_{3,4} < 240$~MeV/$c^2$) where the $\klpgg$ 
signal is smaller. Due to mis-reconstructions
one of the $\pi^0$ may be  reconstructed with a wrong mass. 
This can happen in case of a $\gamma$-conversion or $\pi^0$ 
Dalitz decay with one electron outside the
calorimeter acceptance that biases  the photon energy reconstruction. 
We use the first three drift chambers as a veto against conversions 
and Dalitz decays if there are hits in coincidence, 
$\pm 20$~ns, with the event time.
In addition we reject events with an overflow condition in the readout
of more than one plane of the drift chambers.
This condition is defined by the occurrence of more than seven hits 
in a  DCH plane in any 100 ns time interval.
In this case the front-end readout buffers are reset
and the DCH information cannot be used reliably. \\
The remaining $2\pi^0$ background is reduced requiring 
$p_{T,4}\,>\, 40$~MeV/$c$, where $p_{T,4}$ is the transverse momentum
to the kaon direction  
of $\gamma_4$, which is the lowest-energy photon of the non-$\pi^0$
pair of photons.
The variable $p_{T,4}$ is on average smaller for $2\pi^0$ events than 
for $\pgg$ events and even lower when  the photon energy is
underestimated. 
Although this requirement reduces the acceptance for 
$\pgg$ at high $y$ and low $m_{3,4}$, it does not degrade the sensitivity 
to the $B$ amplitude as the $\left(y^2-y_{\rm max}^2\right)$ 
factor in equation \ref{eq:ampli} is already small.\\
\subsection{\bf $3\pi^0$ rejection.} 
\begin{figure}[ht]
\begin{center}\mbox{\epsfig{file=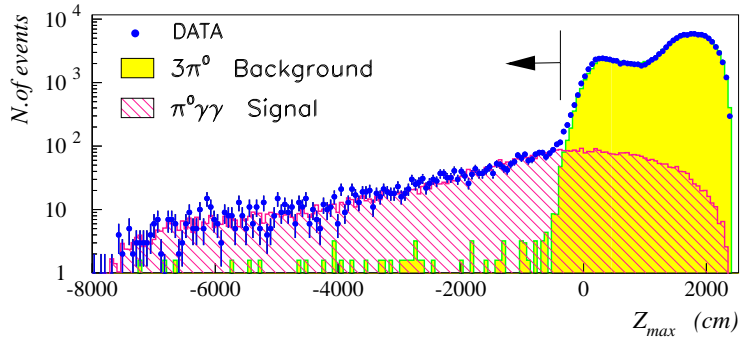,width=10cm}}\end{center}
\caption{\it Distribution of the variable $Z_{max}$ used to reject 
the 3$\pi^0$ background with missing photons, as it looks after all the
other requirements are applied. Dots are $\pgg$ candidates and solid
line Monte Carlo simulation.
The arrow indicates the selected $Z_{max}$ region.}
\label{fig:Zmax}
\end{figure}

In order to reduce the background from $\klppp$ decays,
we require the event to have only four clusters in-time, within 3~ns,
and no activity in the AKL veto counters in coincidence with the event.
As most of the $3\pi^0$ background 
events have one or two photons outside the LKr acceptance, the
reconstructed  $z_K$ 
according to formula (\ref{zk}) moves downstream the beam axis by at
least 6 metres. 
Therefore, the fiducial volume is restricted to the first $30$~metres 
of the $K_L$ decay region.
Events with missing energy are also reduced  by requiring the 
center of gravity, computed from the first momenta of the clusters ($R_{COG}$), to be less than $4$~cm from the beam axis.\\
The simulation of $\approx 3\times 10^9~\klppp$ events 
shows that after applying these cuts the background is composed of events with missed photons:
$36\%$ of the sample has two missing photons, 
while $61\%$ of the sample results from one 
photon missing the detector and two showers overlapping in the LKr. 
The remaining events have all the six photons inside 
the calorimeter but only four reconstructed because of two overlaps.
The corresponding showers are broader than for a single 
isolated photon.\\  
Events with missed photons are rejected on the basis of a variable,
$Z_{max}$, designed to estimate, for these events, the true kaon decay vertex
position. The same estimate, when
applied to signal events, produces an unphysical result, with a
distribution showing many events in the region upstream of the final
collimator. The distribution of this variable is shown in 
figure \ref{fig:Zmax}.
A condition $Z_{max}\leq -5$~m rejects almost $99\%$ of
the $3\pi^0$ background while retaining $46\%$ of the signal.\\
The procedure to define $Z_{max}$ is based on the analysis of the possible
topologies of events with missed photons, as shown in the following
figure:
\begin{figure}[ht]
\begin{center}\mbox{\epsfig{file=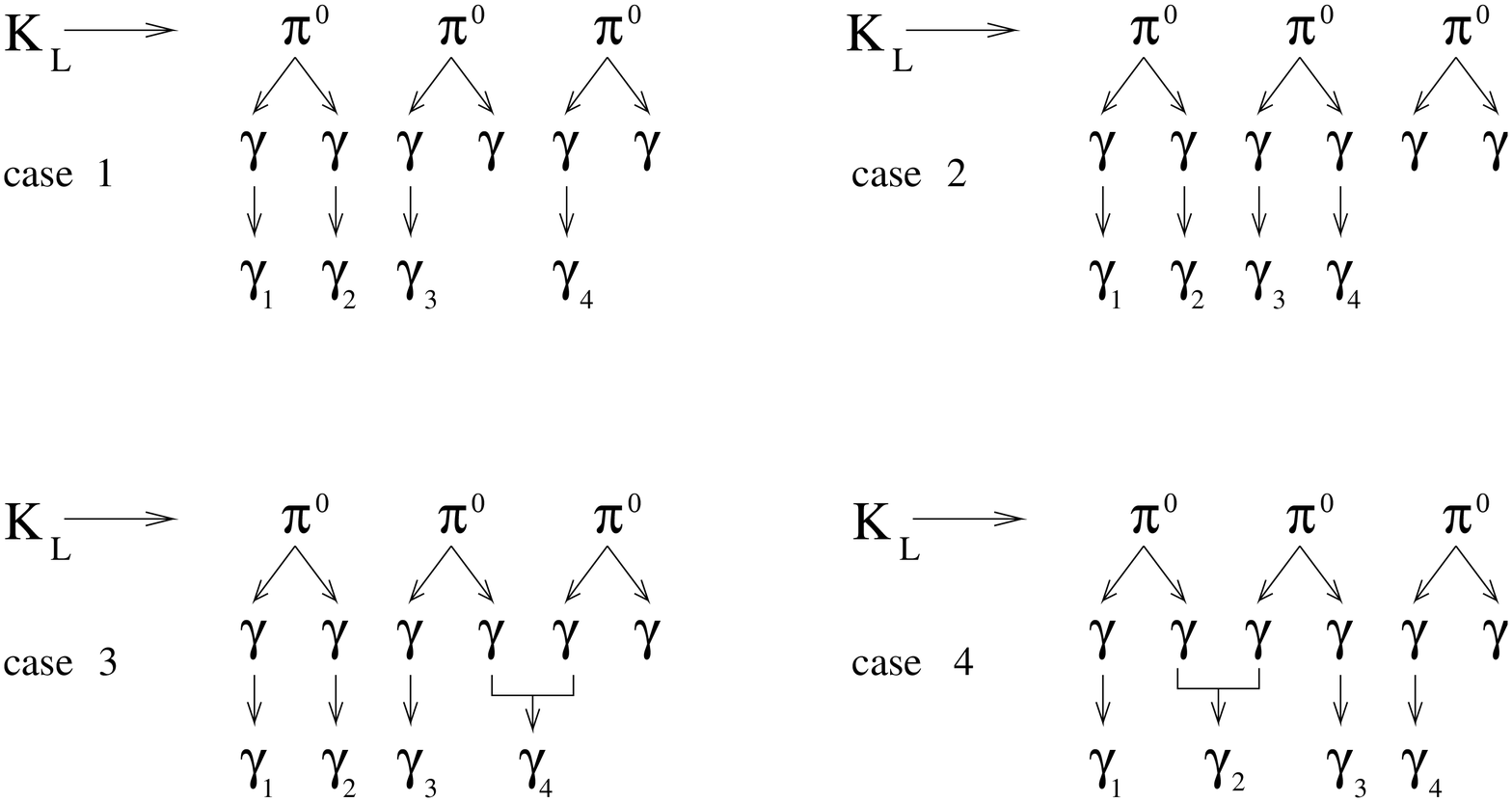,width=13cm}}\end{center}
\label{fig:casi}
\end{figure}
\bei
\item[]  $\underline{\rm\bf cases~~1,~2~~and~~3:}$\\
a $\pi^0$ can be reconstructed and the true vertex coordinate 
$z_{\gamma\gamma}$ is obtained using equation (\ref{mgg}) with 
$m_{\gamma_i \gamma_j}~=~m_{\pi^0}$.
There are five possible $\gamma\gamma$ combinations, other than the
one of the reconstructed $\pi^0$, but we consider 
only those that have $z_{\gamma\gamma}$  smaller than ($z_K - 6$~m). 
We call $Z_{\pi}$ the vertex estimate which is closest 
to this limit.
\item[] $\underline{\rm\bf case~~4:}$\\
one $\pi^0$ has a missing photon and two photons from the
other $\pi^0$'s overlap. Starting from the hypothesys of two $\pi^0$s 
which share a photon, we compute, for each group of three photons, 
the vertex coordinate $z_{\gamma\gamma\gamma}$, assuming each 
photon in turn to be the superposition of two photons  with 
the same position in the calorimeter.
For example if photons 1, 2 and 3 come from two $\pi^0$s and photon 2 
is overlapped, then $z_{\gamma\gamma\gamma}$ is computed as: 
\beq       
z_{\gamma\gamma\gamma}= z_{LKr} - \frac{1}{m_{\pi^0}} \left(\frac{1}{E_1 E_2 d_{1,2}^2}+
                                         \frac{1}{E_2 E_3
                                           d_{2,3}^2}\right
                                           )^{-\frac{1}{2}} 
\eeq
Out of the twelve possible combinations we consider only the triplets
with an invariant mass below $(m_K-m_{\pi^0})$ and the vertex estimate 
closest to, but smaller than, $(z_K-6~m)$.
\eei
For each event we define the variable
$Z_{max}=max(Z_{\gamma\gamma},Z_{\gamma\gamma\gamma})$. 
The residual $3\pi^0$ background originates from two overlaps of
photons such that the $z_K$ is reconstructed unbiased, 
but the shape of one or two showers
is different from that of a single isolated photon. As estimator of the cluster
width we use the RMS of its energy profile in $x$ and $y$:
\beq
 w_{x(y)} = \sqrt{\frac{\sum_i E_i\cdot \Delta x(y)_i^2}{\sum_i E_i}}
~~~~~i\in {5\times 5 \rm ~~cells~around~the~impact~cell}
\eeq
where $\Delta x(y)_i$ is the distance in the $x(y)$ direction of the
$i$-th cell from the center of gravity of the cluster energy.
An energy-dependent cut is applied to the width of each cluster.
The value of the cut is defined in terms of the average $\overline{w}$ and the
standard deviation $\sigma$ of the distribution of $w$ measured in
photons from good $2\pi^0$. We require $w < (\overline{w}+1.8\sigma)$ on the
photons of $m_{3,4}$ and $w < (\overline{w}+3\sigma)$ on the
photons of $m_{1,2}$.
The signal reduction is at the level of $15\%$.

\section{\bf Background evaluation}
\begin{figure}[ht]
\begin{center}\mbox{\epsfig{file=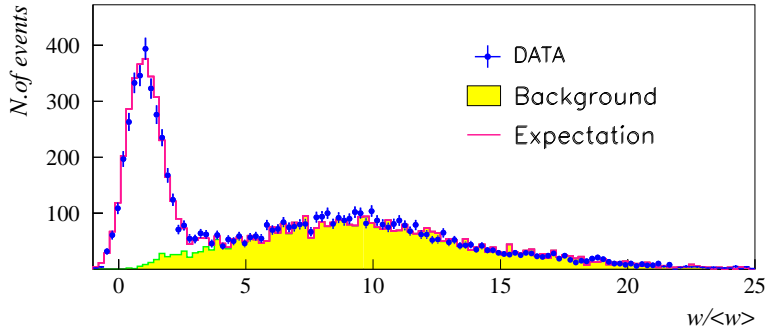,width=11cm}}\end{center}
\caption{\it Distribution of the $w$ variable used to reject the 3$\pi^0$ 
background (shaded area) with overlapped showers after all the other requirements. 
The white histogram corresponds to the sum of signal and background 
distributions. 
Monte Carlo data (solid line histogram) follow the experimental 
data (dots).}
\label{fig:rms}
\end{figure}

\begin{figure}[ht]
 \begin{minipage}[t][6cm][t]{0.45\linewidth}
 \centerline{\epsfig{file=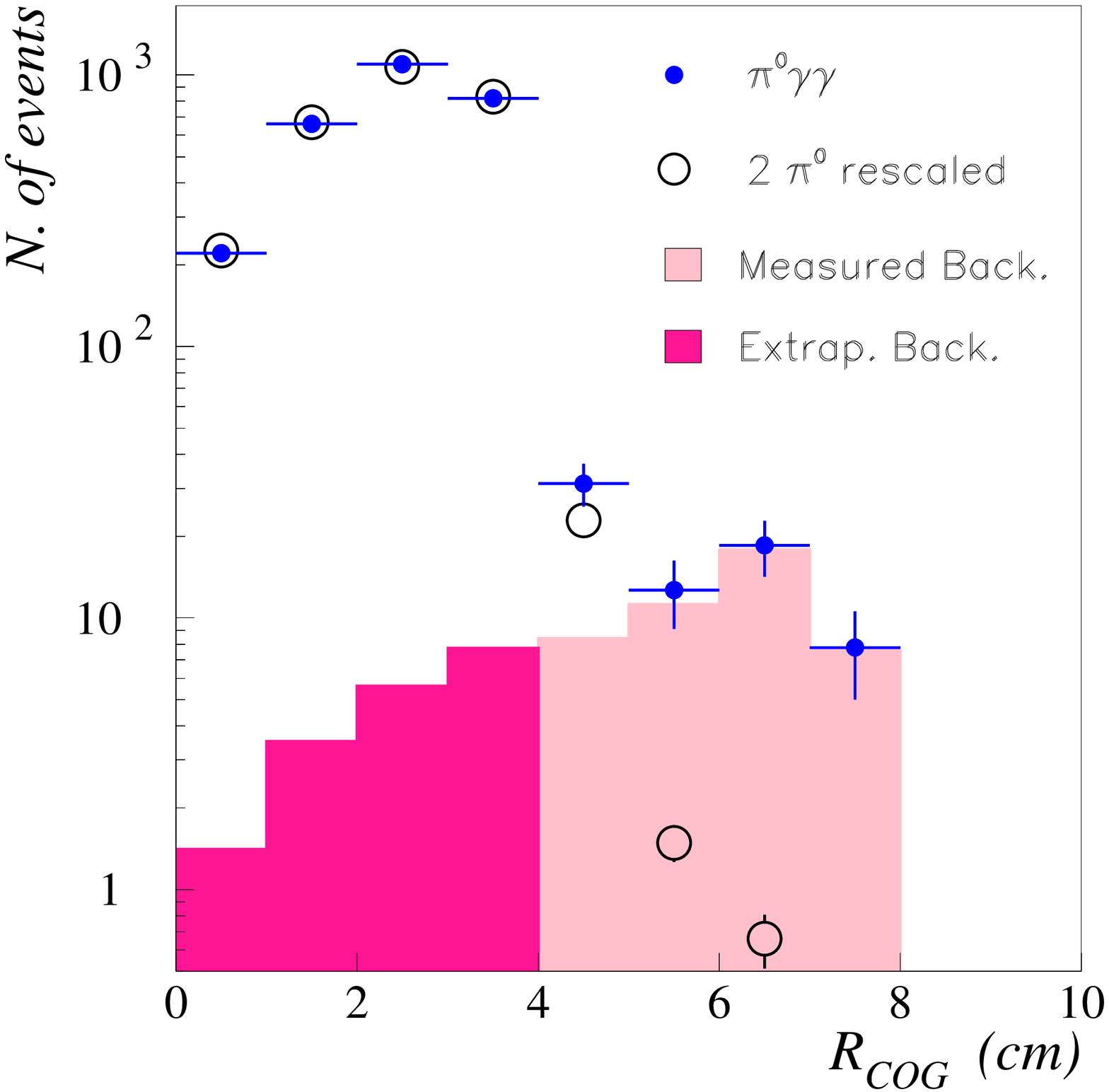, width=6.5cm, height=6.5cm}}
 \end{minipage} \hfill
 \begin{minipage}[t][6cm][t]{0.45\linewidth}
 \centerline{\epsfig{file=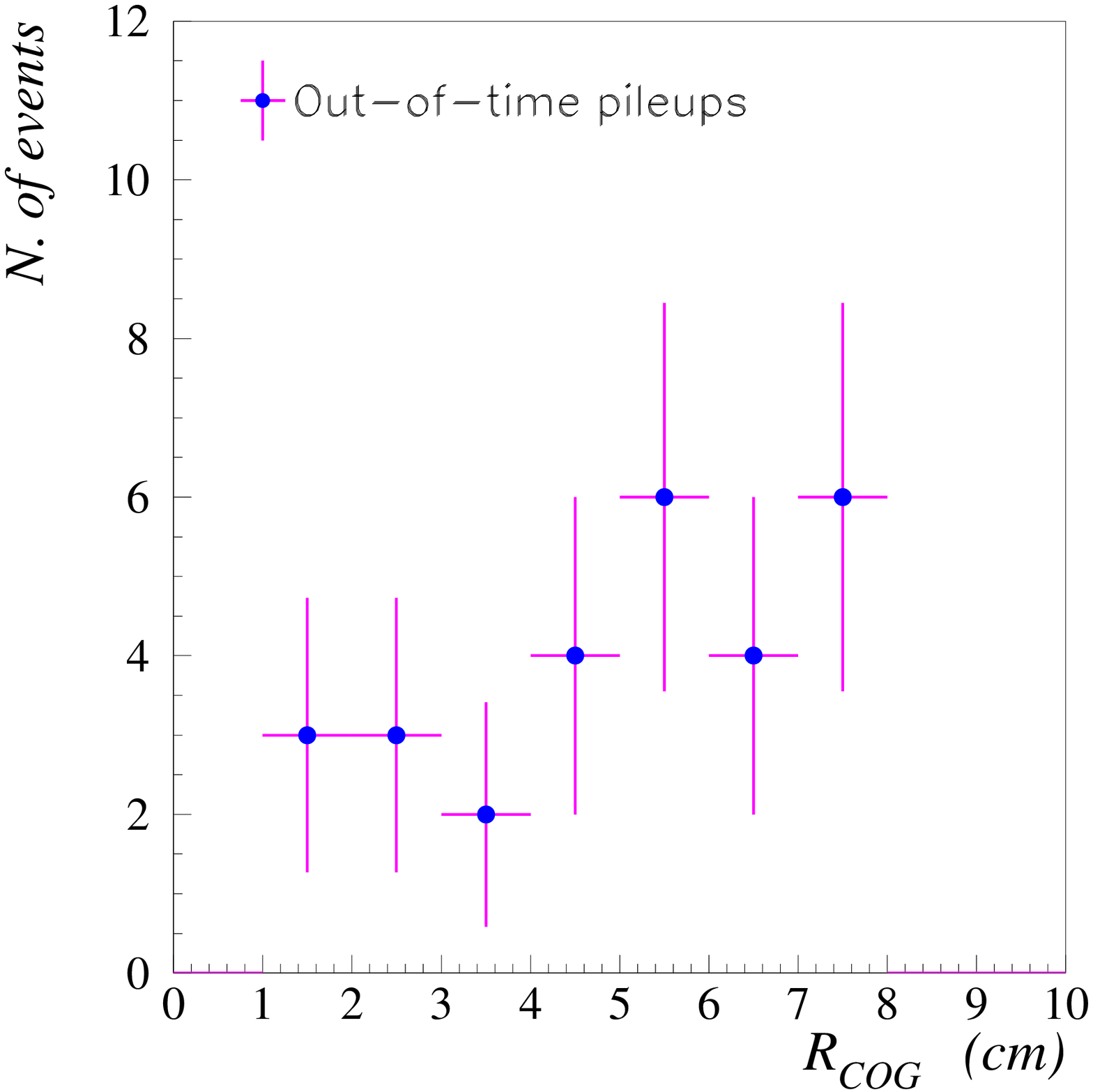, width=6.5cm, height=6.5cm}}
 \end{minipage} \hfill 
\caption{\it $R_{COG}$ distribution (left) for the candidate events (dots)
for 2$\pi^0$ events (open circle). 
The residual background in the tail of high $R_{COG}$ is linearly extrapolated
below the peak as suggested from the $R_{COG}$ distribution of events with 
out-of-time clusters (right).}
\label{fig:cogback}
\end{figure}

\begin{figure}[ht]
 \begin{minipage}[t][6cm][t]{0.45\linewidth}
 \centerline{\epsfig{file=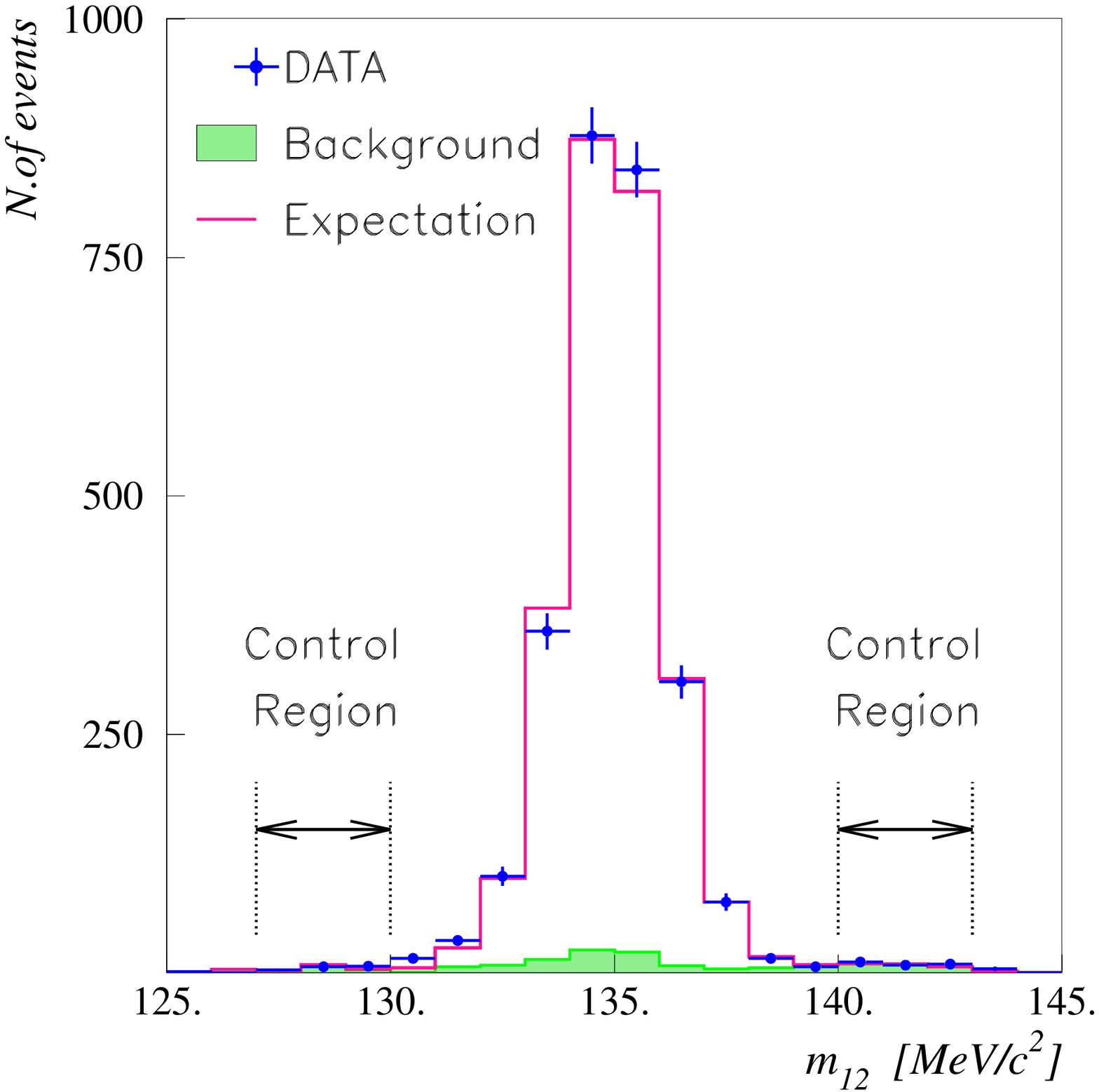, width=6.5cm, height=6.5cm}}
 \end{minipage} \hfill
 \begin{minipage}[t][6cm][t]{0.45\linewidth}
 \centerline{\epsfig{file=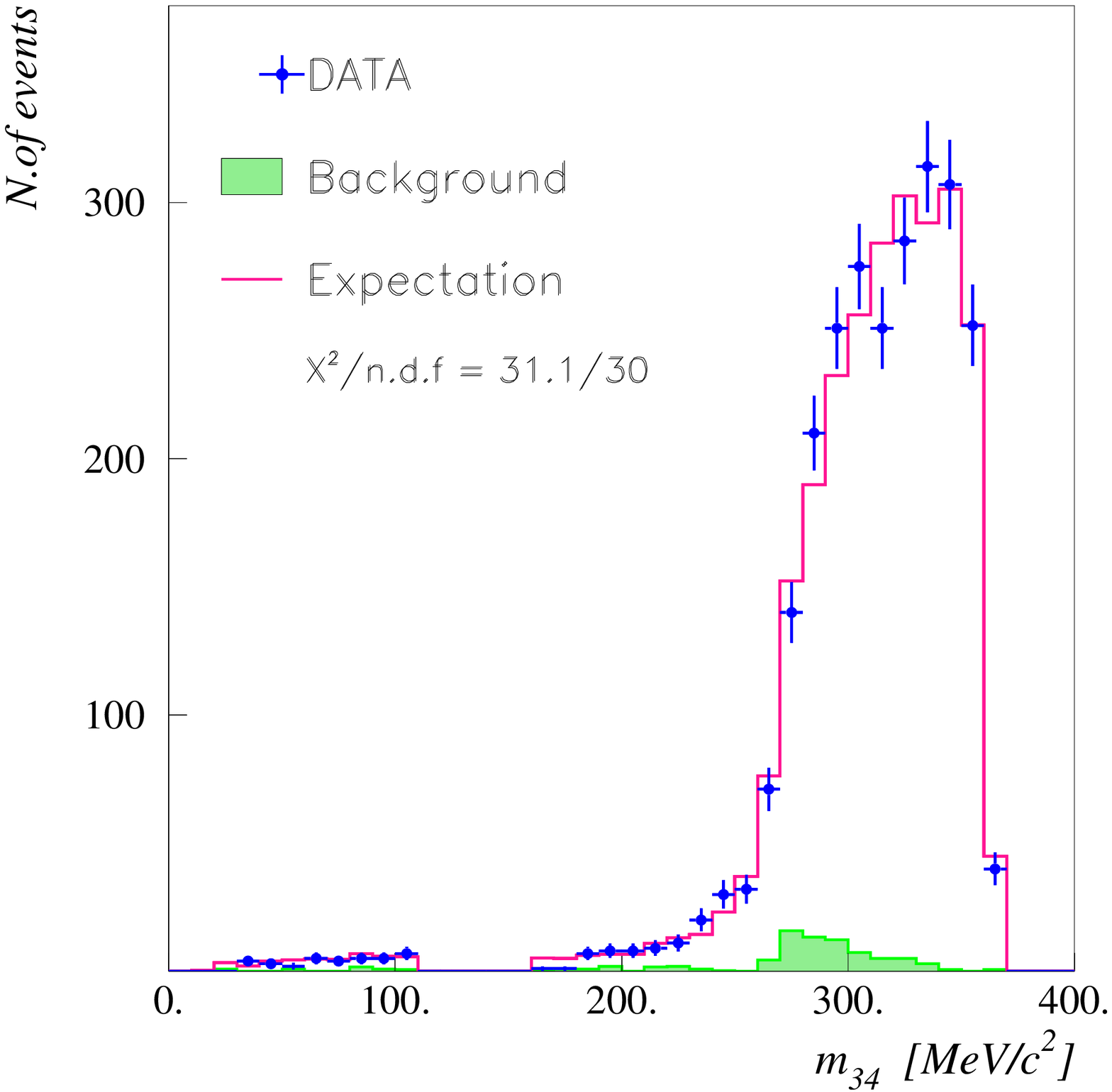, width=6.5cm, height=6.5cm}}
 \end{minipage} \hfill
\caption{\it $m_{1,2}$ and $m_{3,4}$ mass distributions for the $\klpgg$ 
candidates events (dots) and Monte Carlo data (solid histogram) where the 
fitted value of $a_v$ has been used in the event generation. The shaded 
histogram is the sum of the three background types. In case of
ambiguous mass assignment, only the combination with $m_{1,2}$ closest
to $m_{\pi^0}$ mass is included in these distributions.}
\label{fig:m12m34allback}
\end{figure}

The $\klpgg$ candidates that survive all the  above selections are
2558 in the signal region defined as $132< m_{1,2}<
138$~MeV/$c^2$. For 345 of these events more than one combination
satisfies the $m_{1,2}$ cut and therefore the mass assignment is ambiguous.\\
We estimate the contamination from $2\pi^0$ decays and pile-up
processes directly from the data. To quantify the
$K_L\rightarrow \pi^0\pi^0\pi^0$ background we rely on Monte Carlo.\\
We study the $\klpp$ background in the signal region using $\kspp$
decays from different $K_S$ data samples: the proton tagged $K_S$ sample, collected
concurrently with the $K_L$ data, and a special $K_S$ high intensity run ($K_SHI$) taken in 1999.
In the case of the $K_S$ tagged sample we ask for a tight coincidence
within $\pm 0.5$~ns between the tagger and the LKr signals. 
In this way we enrich the number of  $2\pi^0$ events contributing to
the background. Conversely the only contribution to $\pgg$ in this
sample comes from the $K_L$ contamination due to accidental
coincidences in  the tagger, which is estimated to be $3.5\%$ with the
cut used.
\begin{figure}[ht]
\begin{center}\mbox{\epsfig{file=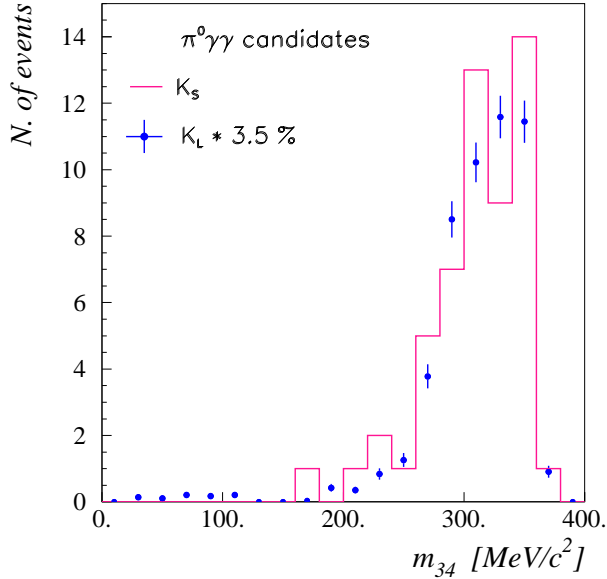,width=8cm}}\end{center}
\caption{\it $m_{3,4}$ distribution for the  $3.5\%$ of the $K_L$ data 
(dots) and for the candidates found in the $K_S$ tagged sample (histogram).  
 No excess is visible in the low-mass region}
\label{fig:2pi0back}
\end{figure}
After applying the $\pgg$ selection criteria, the 
$m_{3,4}$ invariant mass distribution is shown in
figure \ref{fig:2pi0back} for events in the tagged sample 
compared to the original $\klpgg$ mass distribution scaled by the 
$K_L$ contamination factor.
The $2\pi^0$ background is estimated from the tiny excess of tagged
events measured where the $\pi^0\gamma\gamma$ signal is smaller,
namely for $m_{3,4}\!<\!0.240$~MeV/$c^2$ while for
$m_{3,4}\!>\!0.240$~MeV/$c^2$ it is linearly estrapolated from 
control regions on the sidebands of the $\pi^0$ peak 
( $127< m_{1,2}< 130$~MeV/$c^2$ or $140< m_{1,2}< 143$~MeV/$c^2$).
This procedure leads to a background evaluation of  
$(4.1\pm 1.6_{(stat)} \pm 1.4_{(syst)})$~events.
Here the systematic error is related to the 
subtraction of the $K_L$ contamination and to the fact that $K_S$
events have a distribution of the decay vertex  populated
only in the region between 6 and 18 metres.\\
The high intensity $K_S\,$ data sample provides about $60$ \% of the 
tagged $\kspp$ statistic with negligible $K_L$ component.
The number of background events found 
is in agreement with the previous estimate.\\
The cuts applied to reduce the $\klppp$ background are mainly of
geometrical nature and can safely be simulated. The most critical cut 
is the one on the cluster width, for which an accurate description of the LKr 
response for overlapping showers is needed.
For this last variable we  compare data and Monte Carlo distributions 
in figure \ref{fig:rms}.
The simulation describes the data within $3\%$ both in shape and
quantity being normalised to the  Kaon flux.
After the cut the estimated residual background is $(70.2\pm 8.2_{(stat)}\pm 
7.0_{(syst)})$ events, where the systematic uncertainty comes from the non-Gaussian 
tail of the calorimeter response mainly due to photo-hadron reactions.\\
Finally, the pile-up events, which could originate for example from
two overlapped in-time kaon decays, can be quantified comparing the normalised 
distributions of the center of gravity at the LKr for $\klpgg$ candidates 
and for a good $2\pi^0$ data sample. 
An excess at $R_{COG}> 4$~cm is found in the $\klpgg$ distribution, 
after having subtracted the $3\pi^0$ background (figure \ref{fig:cogback}).
When the excess is linearly extrapolated under the $\klpgg$ signal 
we obtain a pile-up background level of $(8.1\pm 3.5_{(stat)}\pm
4.0_{(syst)})$ events.
The choice of extrapolation function is dominating the systematic
error; however the linear one is expected from a uniform center of
gravity  distribution and is confirmed by the $R_{COG}$ distribution 
for events with out-of-time clusters.\\
In Table \ref{tab:events} is reported the summary of the background
evaluation in the signal region and, as
cross-check, in 
the control regions defined above.\\
\begin{table}[ht]
\begin{center}
\begin{tabular}{l|c|ccc|cc}
\hline
\     sample            & $\pi^0\gamma\gamma$ & $3\pi^0$ & $2\pi^0$ &
Pile-up & Total bck  & \%
  \\
\hline
 signal     & 2558 &70.2$\pm$10.7&4.1$\pm$2.1&8.1$\pm$5.3&82.4$\pm$12.1& 3.2   \\
 sidebands  &   44 &30.6$\pm$ 5.9&2.1$\pm$1.6&4.8$\pm$3.5&37.5$\pm$ 7.0& -   \\
\hline
\end{tabular}
\vspace*{0.3cm}
\caption{\it Number of candidate and background events in the signal
  region ($132< m_{1,2}< 138$~MeV/$c^2$) and in control regions 
($127< m_{1,2}< 130$~MeV/$c^2$ or $140< m_{1,2}<143$~MeV/$c^2$) on the
sidebands of the $\pi^0$ peak.}
\label{tab:events}
\end{center}
\end{table}
The overall background is  $3.2\%$. The invariant mass distributions
$m_{1,2}$ and $m_{3,4}$ are shown in figure \ref{fig:m12m34allback}.
The background estimate is in agreement with that found  in the two control 
regions where 44 background events are observed in the data with an expectation of
$38\pm 7$.

\section{\bf Results and discussion}

\begin{figure}[ht]
\begin{center}\mbox{\epsfig{file=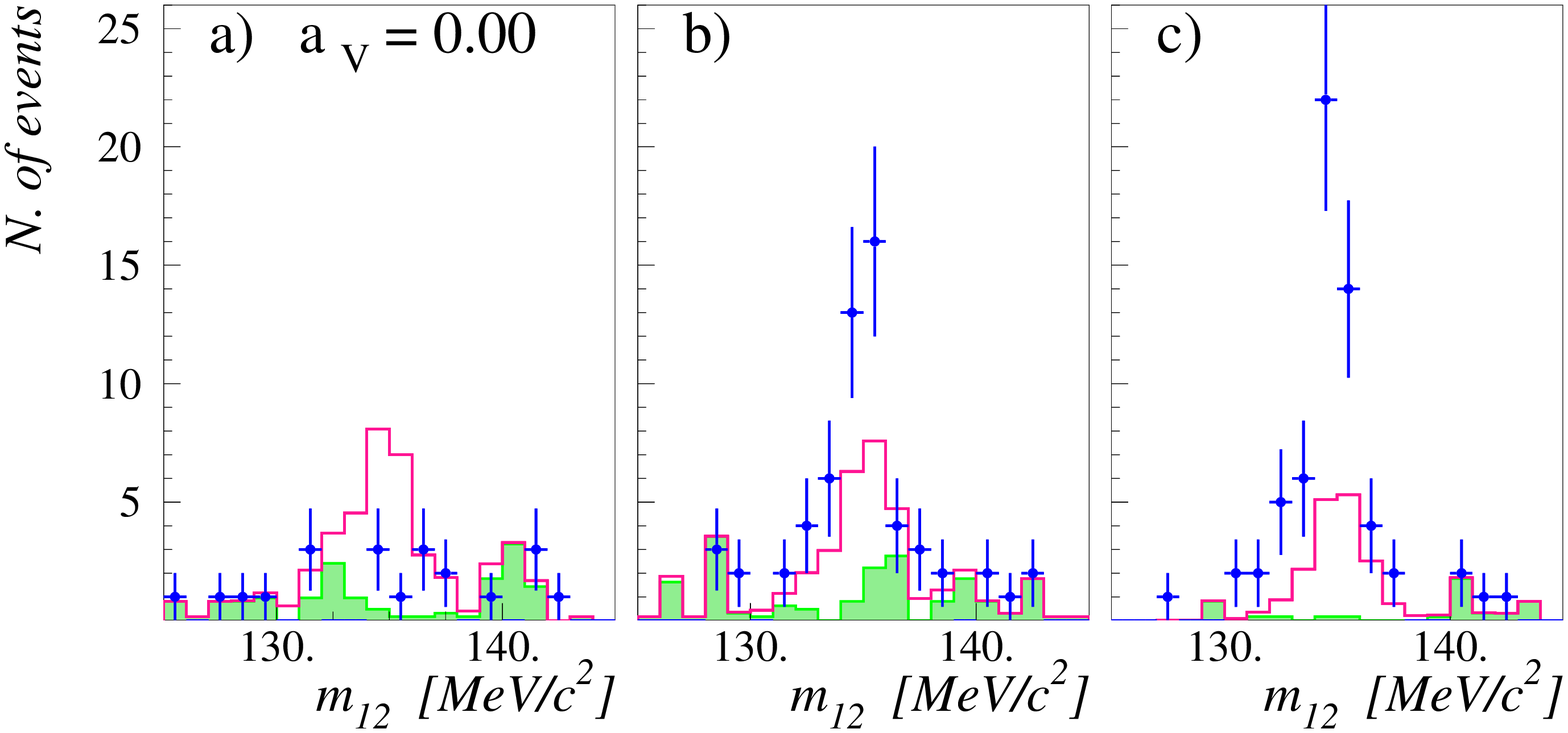, width=13.0cm, height=5.0cm}}
\end{center}
\begin{center}\mbox{\epsfig{file=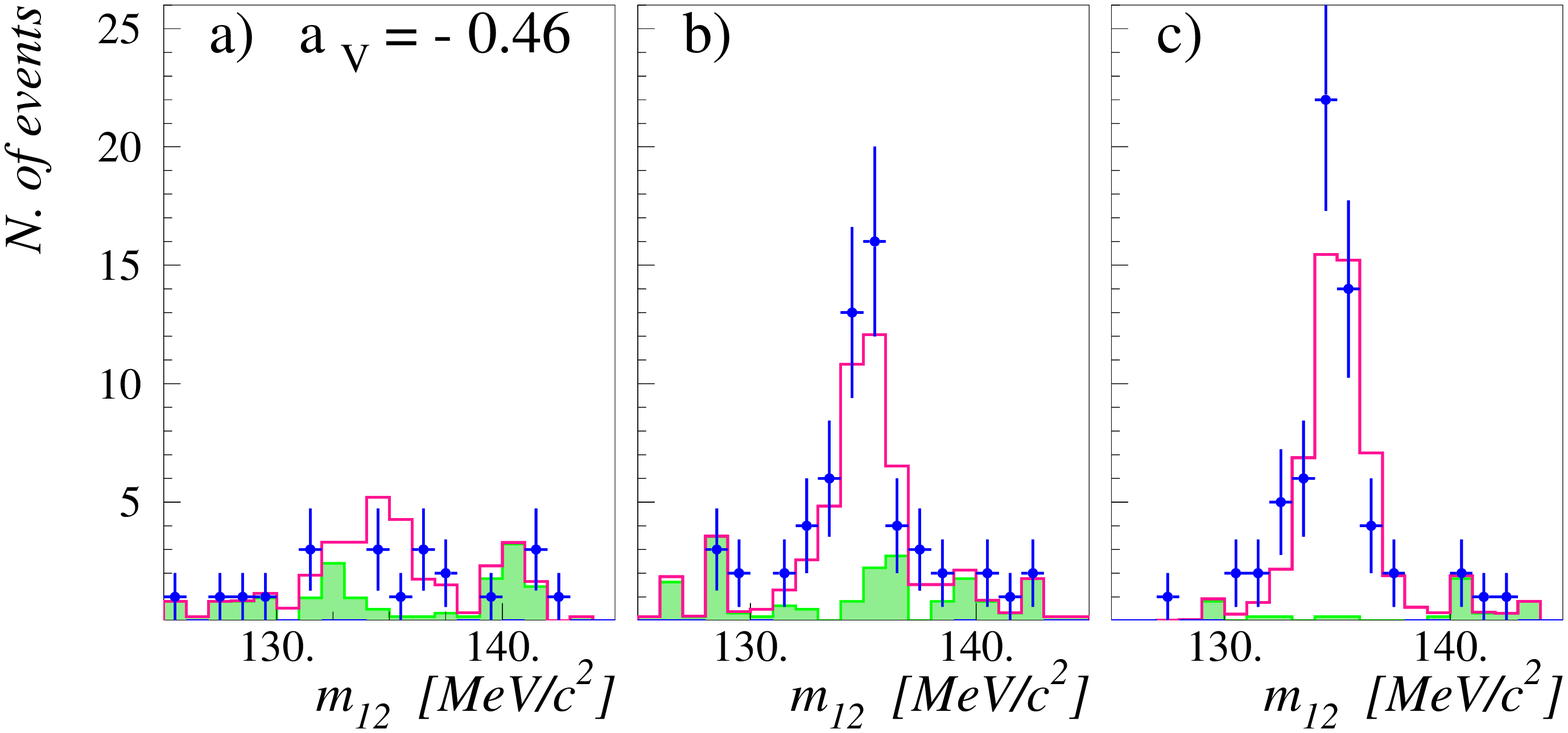, width=13.0cm, height=5.0cm}}
\end{center}
\caption{\it Comparison between data (crosses) and the $O(p^6)$ expectations for 
 $a_v=-0.46$ and $a_v=0$ (solid line histogram) in the $m_{1,2}$ 
 variable for the three regions of the low-mass spectrum:
 {\rm a)} $m_{3,4}\in[30-110]$, {\rm b)} $m_{3,4}\in[160-240]$ and 
 {\rm c)} $m_{3,4}\in[240-260]$.  Background is shown in the shaded area histogram.}
\label{fig:av}
\end{figure}

As confirmed by the Monte Carlo, most of the events in the low $m_{3,4}$ mass 
region shown in figure \ref{fig:m12m34allback} are genuine $\klpgg$ events 
which suffer from the wrong combination of photons.
Events with an ambiguous assignment for $m_{1,2}$ have a flat distribution 
in the $m_{3,4}$ mass region which will inflate the amount of events
in the low-mass $m_{3,4}\!<\! 0.240$~Mev/$c^2$ region. 
To extract a value for the vector coupling constant $a_v$ 
we exclude  events with ambiguous mass assignment.
The events in the low-mass region are thus reduced by $\approx 50\%$. 
In Table \ref{tab:fit} we report the numbers of $\pgg$ candidate, 
background and acceptance  in the 20 bins of the $[0-400] \;$
MeV$/c^2$ $m_{3,4}$ interval.
\begin{table} 
\begin{center}
\begin{tabular}{|c|r|c|c|}
\hline
$m_{3,4}$ $[MeV/c^2]$ & $\pgg$ n-a events & background events & acceptance \\
\hline
    $[0-20]$ &   0  &  0.0  & 0.018  \\
   $[20-40]$ &   0  &  1.0  & 0.060  \\
   $[40-60]$ &   1  &  0.8  & 0.076  \\
   $[60-80]$ &   3  &  0.2  & 0.073  \\
  $[80-100]$ &   3  &  1.9  & 0.069  \\
 $[100-110]$ &   2  &  0.6  & 0.062  \\
 $[110-140]$ &   0  &  0.0  & 0.000  \\
 $[140-160]$ &   0  &  0.0  & 0.000  \\
 $[160-180]$ &   2  &  1.3  & 0.041  \\
 $[180-200]$ &   9  &  2.1  & 0.040  \\
 $[200-220]$ &  12  &  1.0  & 0.044  \\
 $[220-240]$ &  23  &  1.9  & 0.054  \\
 $[240-260]$ &  53  &  0.5  & 0.060  \\
 $[260-280]$ & 184  & 16.6  & 0.062  \\
 $[280-300]$ & 405  & 23.9  & 0.065  \\
 $[300-320]$ & 465  & 10.6  & 0.064  \\
 $[320-340]$ & 514  &  4.6  & 0.063  \\
 $[340-360]$ & 498  &  0.6  & 0.077  \\
 $[360-380]$ &  39  &  0.7  & 0.093  \\
 $[380-400]$ &   0  &  0.0  & 0.000  \\
\hline
\end{tabular}
\vspace*{0.3cm}
\caption{\it $\pi^0 \gamma\gamma$ non-ambiguous candidates, number of background
  events and acceptance  in 20
  bins of $m_{3,4}$ in the range of $[0-400]\; MeV/c^2$. 
  The normalisation factor, i.e. number of decays  in the 
  fiducial volume $N(K_L)$, is equal to $23.9\times 10^{9}$. This
  number is based on the number of
  measured $N(K_L\rightarrow\pi^0\pi^0)$ decays.}
\label{tab:fit}
\vspace*{1.0cm}
\end{center} 
\end{table}
We perform a fit of the bidimensional distribution of the two 
relevant kinematic variables $m_{3,4}$ and $y= {|E_3-E_4|}/m_K$, 
defining a likelihood function as follows:
\beq
 ln  L = {\sum_{i,N_{bins}}{[N_i(y,\mgg)\cdot 
lnE_i(y,\mgg)-E_i(y,\mgg)]}}
\eeq
where $N_i(y,\mgg)$ is the measured number of events in the $i$-th bin 
while $E_i(y,\mgg)$ is the corresponding 
expectation value given by the simulated $\klpgg$ plus 
the estimated background.
The model adopted for the simulation of the $\klpgg$ process is
described in \cite{Ecker3} and it is normalised to the total number of
$K_L$ decays in the fiducial volume. The uncertainty of the parameterisation of the 
$K_L\rightarrow 3\pi$ vertex \cite{Kambor2} is included in the
systematic error.
The best fit, with a $\chi^2$ of $31.1$ for  $30$ degrees of
freedom, gives:
$$ a_v= -0.46\pm 0.03_{(stat)}\pm 0.04_{(syst)}
$$
The systematic error includes the following contributions: 0.02 from
the uncertainty in the parameters of the theoretical model, 0.03
from the uncertainty of the acceptance evaluation and 0.02 from
background evaluation and subtraction.\\
In figure \ref{fig:av} we show the $m_{1,2}$ distribution together with 
the evaluated background and the $O(p^6)$ $\chi PT$ 
expectation for the fitted value of the vector meson exchange parameter $a_v$
in the three low-mass regions a) $m_{3,4}\in[30-110]\;$MeV/$c^2$, 
b) $m_{3,4}\in[160-240]\;$MeV/$c^2$ and c) $m_{3,4}\in[240-260]\;$MeV/$c^2$.
Our data clearly disfavour $a_v=0$.\\
The branching ratio is obtained from the number of $\pgg$ candidates,
including those with ambiguous mass assignment, normalising to the
number of $\pi^0 \pi^0$ events observed in the same sample of $K_L$ 
decays and selected by the same cuts, in order to
minimise the uncertainty in the acceptance corrections and is
found to be:
$$ BR(\klpgg)= (1.36\pm 0.03_{(stat)}\pm 0.03_{(syst)}\pm
0.03_{(norm)})\times 10^{-6} 
$$ 
Both the correction due to the Dalitz decay difference between the $2\pi^0$
normalisation mode and the 
$\pi^0\gamma\gamma$ decay and  the one due to the $K_L\!-\!K_S$ 
interference in $\pi^0\pi^0$ decays \cite{na48eprime9899} are taken into account.
The uncertainty related to the experimental knowledge of the  $\klpp$ 
branching ratio \cite{pdg} is quoted separately. 
The computed average value of the acceptance for the $\pi^0\gamma\gamma$
process is $7.62\%$ and for the $\pi^0\pi^0$ is $8.72\%$. 
The uncertainty related to this computation 
is the main source of the $2.5\%$ systematic error in the branching
ratio measurement since, as shown in Table \ref{tab:fit}, it has sharp
variations. The residual background estimation, the
uncertainty in $a_v$ and the exclusion of the ambiguous events each 
gives less than 1\%  variation on the branching ratio value.\\
As shown in figure \ref{fig:av} there is no evidence for a signal 
in the invariant mass region 
$m_{3,4}<m_{\pi^0}$ where we count 9 $\pgg$ candidates and
($4.5\pm2.7$) background events. 
Using events with $y<0.2$ in 5 bins of $m_{3,4}$ between $30$ and
$110$~MeV/$c^2$, we compute  a model-independent upper limit \cite{frequent}:
$$ BR(\klpgg)|_{m_{3,4}\in[30-110]~Mev/c^2, y\in[0-0.2]} \!<\!
 0.6\times 10^{-8}~~~~90\%~~C.L. 
$$
For this calculation we use the acceptance computed from events
generated according to phase space, in a restricted region where
it is almost uniform. 
Given the negligible contribution of the amplitude $A(J=0)$,
our measurement of the vector  coupling constant $a_v$ allows the 
CP-conserving component of the $\klpee$
decay \cite{Donoghue}\cite{Valencia} to be predicted as:
$$BR(\klpee)|_{CP-conserving} = (4.7 ^{+2.2}_{-1.8})\times 10^{-13}$$
This suggests that the CP-violation effects dominate in
the $\klpee$ mechanism.

\section*{ACKNOWLEDGEMENT}
It is a pleasure to thank the technical staff of the participating
laboratories, universities and affiliated computing centres for 
their efforts in the construction of the NA48 apparatus, in the 
operation of the experiment, and in the processing of data.
We also grateful to G.D'Ambrosio and G.Valencia for useful discussions.

\end{document}